# Interfacing with history: Curating with audio augmented objects

Laurence Cliffe

*The University of Nottingham, School of Computer Science, Nottingham, UK*

Mixed Reality Laboratory, The University of Nottingham, Computer Science Building, Wollaton Rd, Lenton, Nottingham NG8 1BB, UK. Email: laurence.cliffe@nottingham.ac.uk

Laurence Cliffe is a researcher and artist based at the University of Nottingham's Mixed Reality Laboratory and Virtual and Immersive Production Studio. He received his PhD from the University of Nottingham in 2021.

# Interfacing with history: Curating with audio augmented objects

This article presents and discusses the results from visitors' interactions with two audio augmented reality experiences containing audio augmented objects; physical, real-world objects to which virtual audio sources have been attached. It then proceeds to discusses the commonly identified themes arising from the observation of visitors' behaviour within these experiences and the analysis of their verbal and written feedback. The curatorial potential of audio augmented objects is discussed and, by way of conclusion, their functionality as interfaces to digital audio archival content is proposed, along with their ability to reframe, re-contextualise and create renewed experiences with existing collections of *silenced* museum exhibits.

## Introduction

The design of the two augmented reality (AR) experiences presented in this article, *The McMichael Experiment* and *The Laboratory of Psychical Research,* were largely motivated by an investigation into the potential for a *silent* or *silenced* museum object to act as an interface for digital audio archival content. Within the gallery or museum, such objects present themselves as *silent* or *silenced* for a number of reasons: health and safety, preservation, technical obsolescence or the cost and complex practicalities involved in the ongoing maintenance of antiquated technologies. Within these contexts, such objects manifest, for example, as a radio that can no longer play music or a musical instrument that can no longer makes a sound.

The potential for an exhibition object to act as an interface to audio archival content was explored in a project presented by sound scholar and cultural historian Karen Bijsterveld (2015) and discussed in detail by Mortensen & Vestergaard (2014) within what they termed a *listening exhibition* curated at the Media Museum in Odense,

Denmark in 2012 titled 'You are what you hear'. Through the implementation of their *Exaudimus* system, Mortensen & Vestergaard proposed a way of exhibiting and interfacing with radio heritage which was enabled by the digitisation of analogue audio archival content by the Danish Broadcasting Corporation. Within this approach, we saw how, through design and embodied visitor interaction, the exhibition demonstrated potential as an accessible and immersive interface to the sound archive itself.

The recent digitisation of existing analogue audio archive content has both highlighted the need for new methods to promote exploration and engagement with this type of audio content, as well as providing the potential opportunity to do so (Mortensen and Vestergaard, 2014; British-Library, 2019).

With millions of physical recordings now digitised, indexed and tagged with descriptive meta data (BBC, 2022; The-British-Library-Board, 2022) opportunities for extending these recordings' ability to engage with a wider public through the use of innovative and creative digital-based solutions deserve further exploration (Mortensen and Vestergaard, 2014; Cliffe *et al.*, 2019, 2020; Hjortkjær, 2019).

Although Mortensen & Vestergaard (2014) reported some success in generating engagement with the audio archive content contained within the exhibition, significant issues arose around initiating interaction with, and triggering the playback of the audio archive content. The authors attribute this to what they termed *cultural constraints*, the reluctance of visitors to touch, pick up and directly interact with physical objects within a gallery environment, something which goes against normal behaviour. Unfortunately, the triggering of archive audio playback was largely dependent on these direct interactions.

Both of the experiences presented and discussed within this article utilised mobile AR technology to track users' locations within the physical space of the gallery, and to place virtual sound sources at specific locations within the physical space of the gallery. Despite the deployment of similar technological solutions within exhibition spaces over recent times, the function of this technology has largely been to provide a locationally aware audio accompaniment to the exhibition for the visitor; a high-technological variant of the traditional museum audio guide, where a visitor's location within the gallery environment automatically triggers the playback of associated audio content on their headphones (Bailey, Broackes and Visscher, 2019; Krzyzaniak, Frohlich and Jackson, 2019). Whilst recognising the important capability of such experiences to provide personal and locationally responsive audible contextualisation of exhibition space, the application of this type of technology for the purpose of *reanimating silent* or *silenced* objects within the museum or gallery remains largely unexplored.

In summary, unlike existing locationally aware audio guides, the experiences presented here attach virtual, binaurally rendered audio to *plausible* sound making objects such as gramophones, radios and musical instruments. It is through this approach that *silent* and *silenced* museum objects are *reanimated* through the use of audio archival content that is contemporaneous to the object in question, creating the perception that it is the object itself from which the virtual audio content is emanating. By doing this, exhibition objects are realised that have the ability to both exhibit sound and use sound as a means to exhibit themselves, while maintaining the personalised and embodied, non-direct interactional affordances of modern locationally aware audio guides.

**The Technology**

A smartphone app was developed that delivered binaurally rendered stereo audio to users. Binaurally rendered audio emulates the difference in time and frequency a sound is perceived by the listener's two ears in order to create the realistic perception of a virtual sound being located within a specific position in physical space. Delivered to the listener over a standard pair of stereo headphones, the locative and positional capabilities of mobile AR technology enable this binaurally rendered audio to be perceived in the same way sound sources are perceived in reality (with rotational and translational interactivity, or *six-degrees-of-freedom*); the virtual sounds stay located at their original locations as the listener rotates, and attenuate as the distance between the listener and the sound source increases. Key to this is mobile AR technology's ability to map users' physical environments, track users' movements within this physical environment and accurately place virtual content at specific locations within this physical environment. Using this technology, it is possible to realise audio augmented objects within the museum space by placing virtual sound sources at the precise location of physical objects. If those objects are *plausible* sound making objects, then it is possible that they can be perceived as the physical source of the virtual audio content that the listener is, in fact, hearing through their headphones. The mobile application was designed with technical and economical accessibility in mind, and was capable of operating on an iPhone 6 model (or later) with an ordinary pair of stereo headphones attached.

**The McMichael Experiment**

*The McMichael Experiment* was an AR experience developed and deployed at the National Science and Media Museum (NSMM) in Bradford, UK.

The experience used AR technology to attach virtual sound sources to a vintage radio receiver from the museum's collection. The audio content consisted of archival radio broadcast material, contemporaneous with the chosen radio receiver, along with a contemporary recording of radio static. The experience was designed in a way that enabled participants to virtually *tune-in* to the archival radio broadcasts amongst the sound of radio static by positioning themselves at different points around the radio receiver whilst holding a smartphone running the deployed AR application.

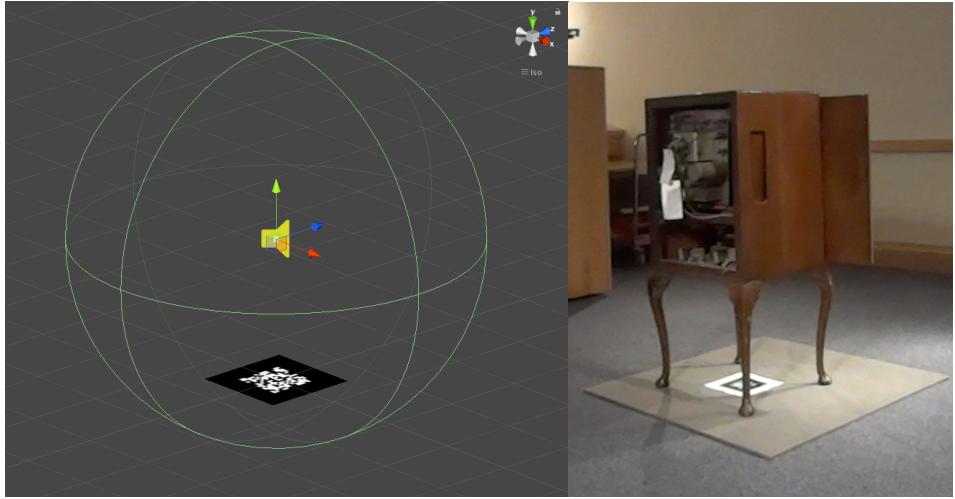

Figure 1. Positioning a virtual audio source in relation to a real-world object using a QR code in the authoring environment (left) and in practice (right).

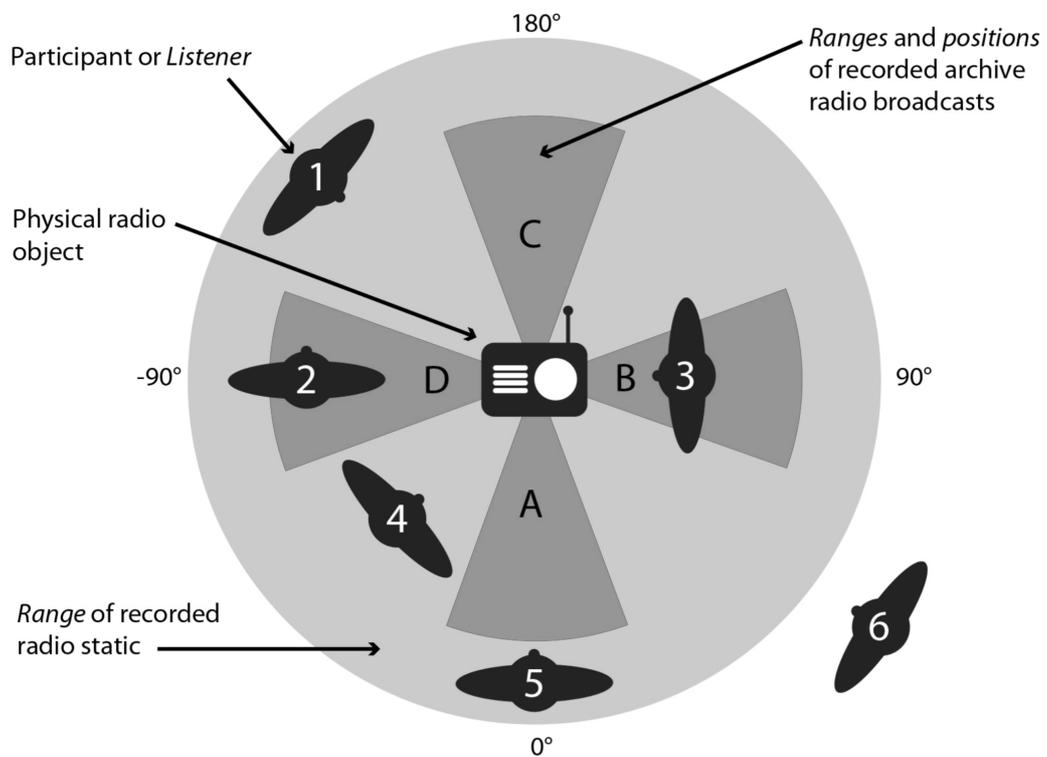

Figure 2. Spatial audio interaction design for The McMichael Experiment.

For *The McMichael Experiment,* a total of 10 participants were provided with an iPhone installed with a copy of the AR application and a pair of stereo headphones. Participants were instructed to explore the radio object and the space around it, no additional details regarding what would happen, how the technology worked or what they could expect were provided.

So that the social interactions between users could be observed and recorded, participants were instructed to undergo the experience in pairs. The decision to study the social interactions between participants stemmed from both an understanding of audio technologies as social technologies (Bull, 2000; Sterne, 2003) and from an understanding of museums as social spaces. Within such spaces not only do visitors have to negotiate and interact with other visitors, but they are often visited by groups, sets of friends and families.

Both video and audio recordings were captured of the participants just prior to, during and after their engagement with the study. Ethical approval was obtained prior to the deployment of the AR experience from the researcher's institution ethics review board, and informed consent relating to the gathering and usage of the recorded data was obtained from all workshop attendees prior to their participation.

Participants were given the opportunity to provide both verbal and written feedback. Verbal feedback was captured on the video camera and took the form of an open-ended discussion. Written feedback was collected on feedback forms; these were completed anonymously by participants and as 'free text' in an effort to encourage the collection of honest thoughts and descriptions of their experiences.

The participants' written feedback was prompted by the question: *How would you describe your experience with the augmented radio?* Verbal feedback was initiated, if participants were not forthcoming on their own accord, by asking what they thought about the experience they had just undertaken. The bodily interactions between all pairs of participants and the radio were recorded on a single, wide-angle video camera that covered the interactional setting of the experience. From this view, participants were recorded entering, interacting with and leaving the setting of the experience.

### *The Laboratory of Psychical Research*

The second AR experience, *The Laboratory of Psychical Research*, ran from the 13th to the 16th of November 2021 within the Court Room of Senate House, University of London as part of the *Being Human Festival*. The experience was situated within a reimaged interior of the historic *National Laboratory of Psychical Research* (1925−1930) founded by renowned paranormal investigator, Harry Price. The exhibition was conceived and curated by Aleksander Kolkowski and kitt price.

As visitors explored the exhibition and scanned the QR codes situated next to the exhibited objects, they built up a virtual soundscape of archival audio content relating to the *silenced* media equipment, such as a wax cylinder Dictaphone, a gramophone and the recording of a séance coming from an empty table and chair. Again, as with the previous experience, these recordings, were delivered binaurally through the visitor's headphones via the mobile AR application.

This binaurally rendered virtual soundscape was accompanied by a distributed speaker-based soundscape designed by sound artist Aleksander Kolkowski, which was carefully designed to both complement and to interrupt, and be interrupted by, the virtual soundscape delivered through the visitors' headphones. With these two soundscapes in play, it was possible to create an immersive and interactive sound experience within the exhibition space where photographs spoke, the recordings of séances lingered over empty tables, percussive instruments randomly sounded and archival interview recordings emanating from antique media playing equipment either abruptly started or became distorted with the sound of static interference as visitors approached. In this way, an interactive mixed reality experience was realised that comprised of physical objects, a distributed speaker system and virtual audio delivered through headphones.

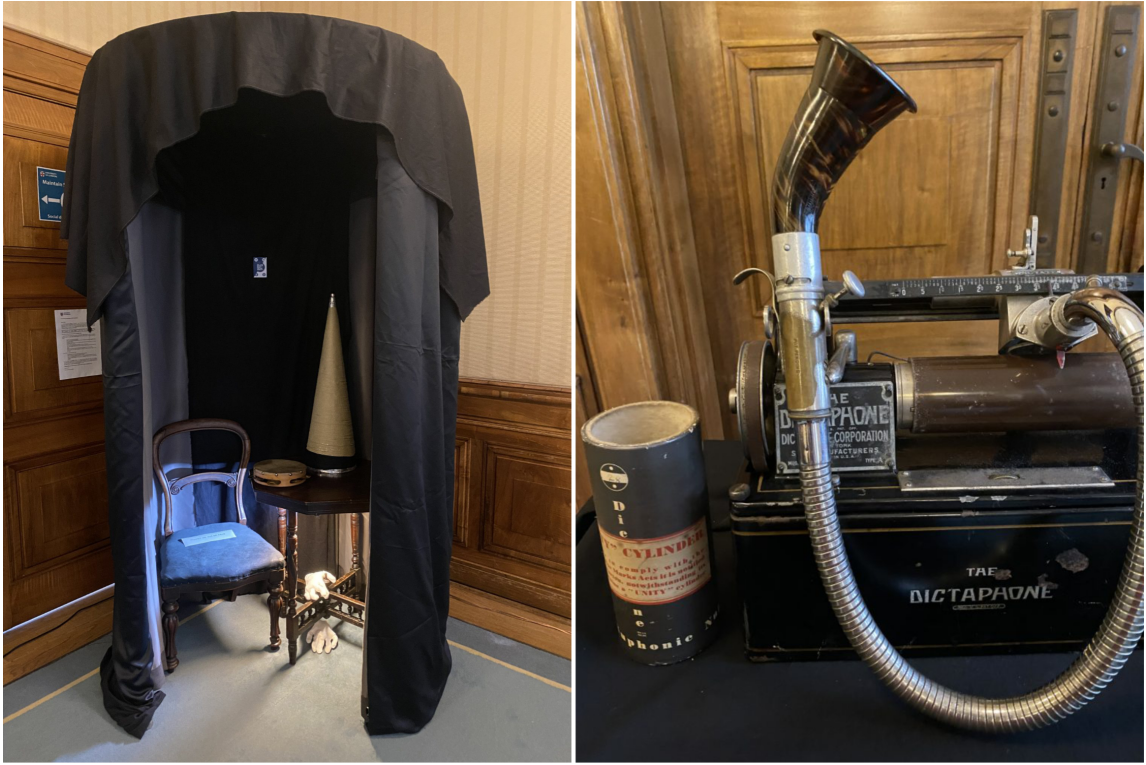

Figure 3. Two audio augmented exhibits from *The Laboratory of Psychical Research*. The séance booth (left) and a wax-cylinder Dictaphone (right).

The mobile application required to experience *The Laboratory of Psychical Research* was published as both an *Android* and an *iOS* application and made available for public download from within the respective app stores. In addition to this, each of the two researchers present each had a phone preloaded with the mobile app that they could lend out to any visitors who did not have a compatible device, or no device at all.

Visitors to *The Laboratory of Psychical Research* sound experience were asked to contribute written feedback regarding their experience, it is these comments that provide the data regarding the visitors' experiences. Over the four-day period 59 visitors to this AR experience left freely written feedback in the comments book.

**Methodology**

The video recordings of participants' interactions within the experience setting, along with their written and recorded verbal comments, act as a resource to facilitate and communicate an understanding of what was taking place within the participants' interactional work. This methodological approach constitutes an ethnomethodologically informed design ethnography, realised by deploying the prototype technology within the setting for which it is intended, and with users for which it is intended, with a view to understanding the human-computer interactions which take place and which can inform subsequent design considerations (Crabtree, Rouncefield and Tolmie, 2012).

A process of interaction analysis was employed to uncover the recurring themes within the participants' activities (Jordan and Henderson, 1995). This analytical process was undertaken inductively, reflecting the six phases of thematic analysis identified by Braun & Clarke (2006) as: data familiarisation, the generation of labels for interesting and recurring activities within the data, the gathering of identified labels under themes, the reviewing of themes, the defining and naming of themes and the reporting on the themes.

**Results**

The following section reports the findings from the two AR experiences in turn.

*The McMichael Experiment*

In the written feedback, all but one of the ten participants described their experience as being either 'interesting' or 'fascinating'. Two participants commented on the authentic

'valve warm sound' and the 'period appropriate programming', one commenting that 'It was interesting to have new technology used to interpret a story about an older object' and that they would like to see this technology used throughout the museum.

Two participants made direct references to how their bodily movements were tuning the radio into the different broadcasts, and likening this to their practical experiences and memories of tuning a traditional radio receiver. There were comments made about being able to listen to individual radio broadcasts, as well as being able to construct or compose a personal soundscape experience from the different elements available; 'picking up and losing the sounds'.

Additional positive references were made to the exploratory nature of the experience and its potential for being adapted as a maze, puzzle or mystery solving experience. One participant mentioned that they would have liked additional visual or textual information displayed on the phone's screen to complement and provide information about the audio they were currently listening to. Furthermore, this feature was suggested as an additional means of navigation within the experience, to visually indicate the whereabouts of specific sounds or, if you miss something, provide a means by which it could be easily found again.

In relation to the verbal feedback, participants identified with the experience of using their proximity and their position in relation to the radio to find the broadcast material amongst the sound of static as being a metaphor for what it may have been like, or what it was like, to originally tune this type of analogue radio receiver, as one participant commented:

> "It reminded me of how difficult and frustrating it used to be to tune a radio, because walking around the object was like tuning it."

Also mentioned, again in relation to the evocation of memory, was the 'Faithful reproduction of the warm valve sound' indicating the potential importance of historical accuracy in the sonic delivery of the audio augmented object. Participants also expressed an interest in further levels of sonic engagement with the object, for example one participant mentioned that they almost expected to hear 'more stations when pointing the phone at the tuning dial on the radio'. Two participants made reference to the 'abstract' nature of the experience and expressed interest in having a more literal and faithful relationship between the object and the delivery of the audio content. One participant commented on how the combination of the real object and the virtual audio triggered their imagination, much like listening to music being a catalyst for the mind's eye, but suggesting that having a physical object in front of them which directly related to the content on their headphones in some way amplified this experience:

> "It just brings the sound out more, so you're kind of just looking at the object, imagining things, the object's actual sounds but without touching it."

We see how, through a process of familiarisation, the participants quickly associate their bodily movements to the receipt of the spatialised audio sources, and then begin to explore the interactional setting to see what they can find. Subsequent to this, we witness our participants returning to investigate the location of some of these sources and engage in listening to them. This phase of *focussed listening* can sometimes result in a more attentive and engaged listening activity, observable by participants attempting to achieve a very close proximity to the location of the virtual sound source. We see how personal space and acceptable social proximities affect the process of virtual sound exploration and investigation, and how these social constraints become more flexible during phases of engaged listening and manufacture an apparent disassociation with the physical environment. We will now look at each of these interactional phase in a bit more detail.

*Preparation*

For the purposes of wider deployment, it is envisaged that the application would be made available for listeners to download onto their own devices, enabling institutions to economically deploy experiences like this, therefore, familiarity and access to an appropriate device would be assumed. Although all participants automatically put on their headphones when they were ready to start, four participants needed to be reminded to put their headphones on the correct way around (essential for the correct orientation of the binaural audio content). Observable from the recorded video of the participants' interactions with the experience, it was noted that two out of the ten participants required instructional prompts to engage in an exploration of the space.

*Familiarisation*

The familiarisation phase is distinguishable within the video recordings by the various lateral movements the participants made. This seems to indicate an initial process of familiarisation with the association between bodily movement and the interactive positioning of the surround sound. These movements are often terminated by an acknowledging sign of appreciation, perhaps confirmation that the association has been recognised, understood and appreciated. These lateral movements were observed being performed in a variety of different ways. Some participants swayed from side-to-side with their device held in alignment with their body and head. One participant waved their device in a lateral motion within a few moments of starting the experience and kept their body stationary whilst doing so. Another participant rotated their upper body in a lateral motion, and therefore also the device they were holding. During this phase of familiarisation, a detachment of the focal gaze from the screen of the device was often observed. In other words, participants, through their particular process of positional familiarisation, were observing the physical object directly, rather than secondarily through the screen of the device.

*Exploration*

After the brief familiarisation phase described above, all the participants can be observed within the video recordings of their interactions walking around the radio a full 360 degrees, often pausing briefly at the locations of the archive audio broadcast content. The direction of exploration, clockwise or counter-clockwise, was most often determined by the first participant to start moving around the object, equally the length of the participants' pauses at the locations of the audio signals were often determined by one participant resuming their exploration around the radio and prompting the other to resume theirs. This behaviour leads to each member of our pair of participants exploring adjacent locations of the sound source, as one member begins to travel to the location of the next broadcast, so does the other member. This type of exploratory behaviour is observed amongst all our participant pairs, though there are some occasional exceptions. These exceptions appear to take place either when one of the participants has become engaged in the next phase of investigatory interaction, or if the participants appear to have a greater degree of social familiarity with each other. The latter is indicated by an observed acknowledgment of each other, an indication of appreciation for the content they have found and a willingness to share that appreciation with the other participant.

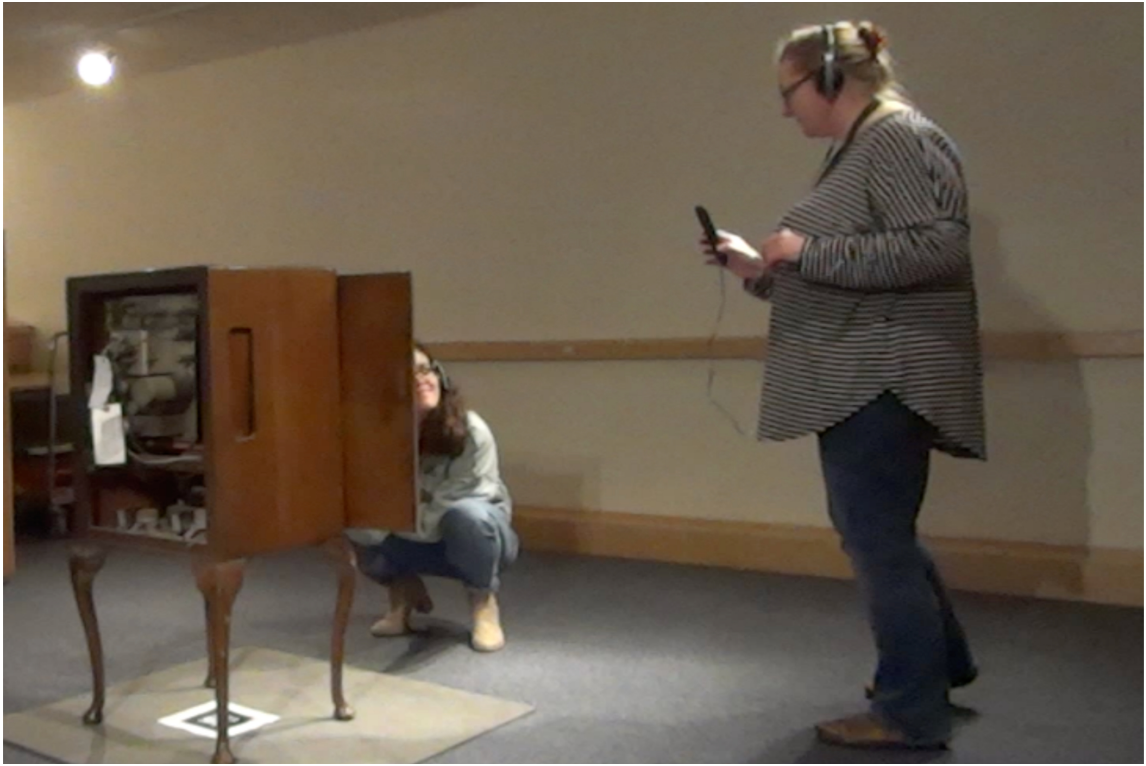

Figure 4. Two participants acknowledging each other's appreciation of the audio content within *The McMichael Experiment.*

*Investigation*

Within this phase, participants are observed returning to the locations of the audio broadcasts that they had initially identified during their exploratory phase. We also begin to see exploratory interpretations of the smartphone device as an interface to the audio content. These interpretations take on a variety of styles, with one participant holding their device aloft in an antennae-type fashion, directly reflecting the subject of both the virtual and the physical, another uses their device as a virtual microphone, moving it towards points of interest around the artefact. Others listen through the window of the screen, or rather, observe the radio through the screen of the device whilst listening through their headphones. During this phase of interactional activity, we also observe participants sharing the same audio sources and interacting in much closer proximity to each other.

*Focussed listening*

The investigation phase, where the participants revisit the virtual audio broadcasts they identified within their exploratory phase, quickly develops into focussed listening. This is discernible within the video recordings of participants remaining stationary for prolonged periods for the first time since beginning the experience. Evident within this interactional phase is an apparent disassociation with the physical object itself, with participants being observed closing their eyes or seemingly focussing on other more distant objects whilst they concentrate on the audio content. This behaviour is also documented in the following participant's written feedback, though it is interesting that despite the visual disassociation with the radio object, a strong sonic and physical attachment to it remains:

> "It was a fascinating experience. The object came alive, I entered a new sonic dimension where I was totally immersed. (I also closed my eyes repeatedly). I was trying to understand the context of sound content, the words of the man speaking."

Again, despite this visual disassociation with the object whilst engaged in these periods of focussed listening, these events initially take place at either the front or the back of the object, areas of distinct visual interest compared with the two rather plane wooden sides, with the exposed electronic and mechanical insides at the rear, and the TV screen and radio dials at the front. This behaviour is observed despite the location of two audio broadcasts at the sides of the object.

*Second-level focussed listening*

Throughout the recordings of all the pairs of participants, it is possible to witness moments when at least one of the participants engages in listening in much closer proximity to the object, often crouching down in order to obtain a physical position very close to the centre of the virtual sound source. Again, this happens exclusively at the front or to the rear of the object where the object's mechanical and electrical interfaces and inner workings can be seen respectively. This observed activity suggests that the visual component of an audio augmented object can form an important part of a user's interactions with the augmented object as well as the audible component.

*Interruption and finishing*

Interruption to a participant's interactions, which often resulted in the termination of their participation, stemmed from one of the pair of participants deciding they had finished. Evident throughout all the recorded interactions, in all but one of the 5 pairs of participants, the end of participation is initiated by one participant removing their headphones, which prompts the other to do the same, even though the participants never started at exactly the same time. In the one event in which this did not happen, the other participant was engaged in *second- level focussed listening*.

**The Laboratory of Psychical Research**

Through the analysis of the written feedback provided by visitors to *The Laboratory of Psychical Research,* we can observe the following recurring themes: *enjoyability and fascination*, *atmosphere and information*, *new gallery experiences*, *a lack of context* and *technical issues*.

*Enjoyability and fascination*

Specifically, within the following participant comments, although also evidenced within subsequent comments, we see how an enjoyable appreciation of the experience is attributed to its immersive quality and to the unique and innovative application of the technology in this context.

> "Great use of technology…"

> "Fascinating immersive experience."

> "Fascinating! Learned a bit and a cool immersive experience."

> "One of the best AR implementations I've seen."

> "A very unique experience. Never been to anything like this before. Sometimes a bit spine-chilling."

Within the following comments, the enjoyment obtained through the participation in the experience is reflected upon further. This time, the enjoyment is attributed to the use of spatial audio and the connection between this, the physical objects, and the speaker distributed audio experience.

> "Amazing links between sounds and object."

> "I really enjoyed the connection of object, place and sound."

> "A fascinating mix of technologies… I enjoyed the mix of spatialised sound and the sound from the objects themselves."

*Atmosphere and information*

Here we begin to unwrap the explicit curatorial potentials of the mixed, audio augmented reality environment. Within the following comments, participants reflect on the experience's atmosphere and its ability to bring the subject to life and to disseminate information and inspire further learning around the subject.

> "Loved the way the audio followed me around the room! Really interesting way to bring alive this subject."

> "An interesting mix of items and sound, brings to life what happened during the research."

> "Evocative combination of technology, historical artifacts and sound."

> "Brilliantly haunting installation – eerie, atmospheric and educational…"

> "Really impressive exhibits, the technical virtuosity of the locators and sound effects can ease you into what you come to realise is an eerie combination of materials."

> "Such an interesting way to put together an exhibition."

> "Wonderful installation, beautifully put together from a technical perspective… I learned and I enjoyed!"

> "Thanks for the installation – I feel like looking into this project."

*New gallery experiences*

In addition to the previously presented comments regarding the perceived uniqueness of the experience, in the following comments participants directly reflect on the potential of this new gallery experience. Whilst one participant reflects upon the directness of the experience, the other comments on the potential for new forms of exhibiting that integrates archival audio content and the creative application of sound.

> "It worked really well with the old artefacts in front of you and the immersive audio recordings. Listening through headphones is also a lot more immediate and direct."

> "Excellent installation – very creative and clearly the beginning of great potential for new forms of display integrating archival material and creative sound composition."

*A lack of information and context*

Four of the participants provide comments regarding the need for more context and information about the exhibition's subject matter. Within the first two comments below we see how there are contrasting views regarding the use of sound, with one participant expecting textual content over what they describe as noise, and the other suggesting that a longer experience of this type could facilitate a greater degree of knowledge exchange.

> "I'd had rather read more about the information. I'd prefer that to random noises."

> "Really enjoyed the installation – would love to attend a longer one with more information. Particularly enjoyed how the sounds came in and out as you move around – very effective!"

> "Interesting immersive experience. I would have liked more context on the subject."

> "I liked the effect once the QR codes had been activated. More intro and context would have been beneficial."

*Technical issues*

Out of the 59 comments that were provided by visitors, 3 describe technical issues relating to installing the mobile application on their phones. These are exclusively the result of a visitor's Android phone not being compatible with *ARCore* via *Google Play Services* for AR, which enables AR experiences on *Android* devices. Though, once working, we can see that initial disappointments and frustrations are often overcome due to engagement with what they described as an interesting, enjoyable, fascinating and atmospheric experience:

"Only one phone in our 5 would run the app."

"Very enjoyable and interesting way to represent the supernatural. Technology was fantastic – which was more than can be said for my old phone, But thank you for lending us yours!"

"Really enjoyed it. Fascinating insight into historic psychical research – very atmospheric - a shame only one of our phones was good enough for the app."

**Discussion**

Within the gallery, a system that relies on the modality of non-tactile, embodied spatial interaction overcomes the problems encountered by Mortensen & Vestergaard relating to their reliance on physical interaction with objects within such an environment (Mortensen and Vestergaard, 2014). Within the gallery, such an approach also enables the explicit authoring and design of attractor sounds (Zimmermann and Lorenz, 2008) as objects need not be physically, or even visually, encountered prior to their presence being made aware to users. This suitability is also reflected by the observed potential of audio augmented objects to engage audiences with both physical objects and virtual audio content, including digital audio archival content.

The thoughts participants shared regarding the potential uses and applications for this type of technology within exhibition spaces, along with their interest in undertaking such an experience, suggests that such an experience would be well received as part of an exhibition. The findings also indicate that the convenience and agency offered museum and gallery attendees through the utilization of their own technology to undertake such an experience is desirable. Specifically, we find participants welcoming the exploratory freedom afforded to them by the model of spatial interaction inherent within the experiences as a refreshing change to the restrictive nature of more widely available audio guides. The enjoyable and performative interactional experiences that this exploratory and spatial freedom facilitates is also worth considering within such a context.

On the point of visitor agency, we can reference Bull's (2013) suggestion that the connectivity to the environment afforded to the listener via the ability to *freely associate* virtual audio and physical reality differs from that which can be achieved via *directed association* (narrative, historical or otherwise). Although Bull (2013) insists that this experience is a product of the nature of the technology, the environment and 'the user's cognitive orientation', the differing that the *agency of association* represents could be one that has the potential to facilitate a greater level of connection with the subject.

Of particular curatorial interest is how the participants' experiences demonstrate the ability of an audio augmented object to communicate its presence, location and character beyond line-of-sight. An audio augmented object's ability to do this can be used to affect the listener's exploration of physical and audible space, by both inviting and deterring exploration. Once seen, both the object's existence and characteristics can remain present for the listener throughout the entirety of the experience regardless of the object's audibility within the soundscape. Furthermore, this sense of presence, along with the characteristics with which it is associated, could possibly extend beyond the scope of the experience. Whilst we can see that the presence of something can be suggested through audio, it's the confirmatory visual association of the virtual audio's source in the guise of a physical object that makes the suggestion believable within the context of the experience, and that this association has a fluidity and malleability that can be creatively exploited.

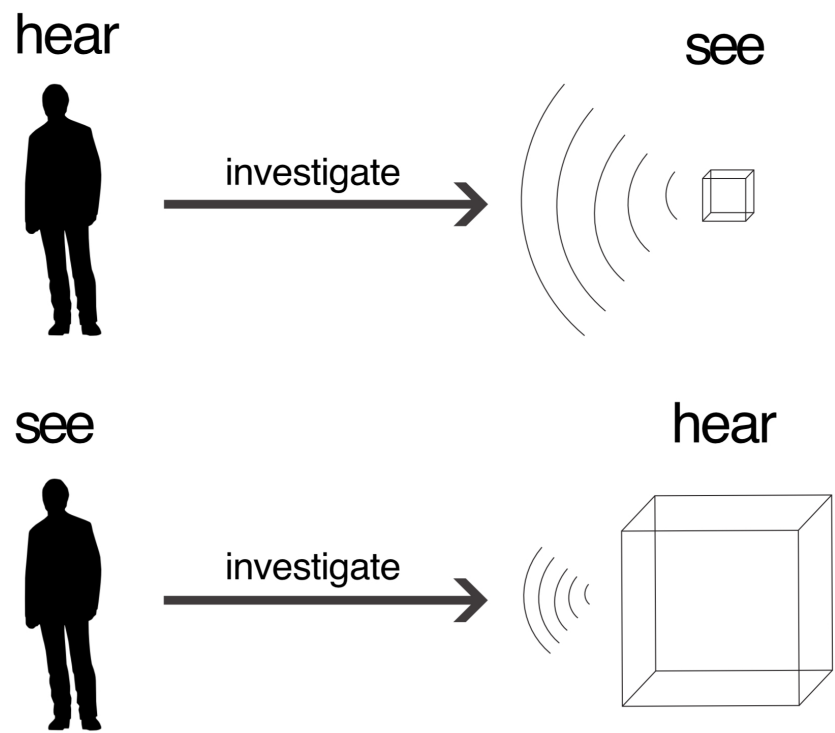

Figure 5. Promoting investigation of the physical object through virtual sound, and investigation of virtual sound through the physical object; the respective audio-first and object-first design approaches.

*An absence of experience*

*Silenced* objects present as objects with which the experience has been lost; the radio that can no longer be tuned, or the musical instrument that can no longer be played. Such objects should not be limited to the media playing or musical type, nor are the sounds of media playing or musical objects limited only to the media content or music that they are capable of playing. Indeed, it is perhaps through a thorough investigation of all possible sounds associated with all sorts of objects that the unusual and the unexpected content can be uncovered that can help entice initial engagement.

Virtual audio, as realised through a museum-based audio augmented reality experience, presents itself as an antidote to this dilemma. It provides a way for visitors to experience objects that cannot be picked up and played, turned on, tuned in and turned up. Of course, whilst not being able to interact with these objects in the usual tactile way, new ways of remotely interacting with them need to be devised but, by thinking carefully about how this can be achieved, not only can an experience with a *silenced* or *silent* object be realised, but an experience with the object's associated audio content can also be attained. Additionally, such interactions have the capacity to help tell the stories and experiences behind the objects, and connect them to relevant and appropriate audio archival content.

By including, or at least considering, all sounds associated with an object as worthy contenders for an object's audio augmentation, it should be possible to curate virtual soundscapes around collections of audio augmented objects that go some way to capturing the experiences and atmospheres associated with specific objects, places and moments in time.

### *Atmosphere as communicative gestalt*

The creation of such a holistic and interpretive sonic experience speaks strongly to the practices of soundscape studies, acoustemology and sonic ethnography (Droumeva, 2016); where the study of a setting's soundscape is attributed with a potential to communicate an in-depth understanding of time and place, more so than the sum of its component sound sources and, at times, to greater effect than perhaps more traditional methods of documentation and enquiry (Lingold, Mueller and Trettien, 2018). Though, perhaps a more considered definition is provided by Gershon (Gershon, 2019), who suggests the epistemological potential of sound, as realised within the practical application of sonic ethnography, lies in its ability to promote an understanding of things from a different, non-ocular point-of-view.

The creation of a thought-provoking sonic atmosphere for exhibiting collections of objects was, to a small degree, realised within the experiences presented here, and seems entirely possible to achieve on a larger scale. This functionality has the potential to be achieved with collections of audio augmented objects which can contribute to an over-arching and communicative sonic atmosphere, as well as affording more detailed investigation of their individual audio content.

A headphone-based binaural listening experience provides an opportunity for further sonic curatorial control and a level of personalisation of the sound experience not possible with a distributed speaker-based solution. With personal headphone audio delivery, the problem of timed and looped audio events is overcome by ensuring that, for example, narrative-based content always begins when the visitor arrives, rather than being half way through its timed or looped cycle. Additionally, different layers of sonic engagement can be delivered simultaneously to different visitors, whilst one visitor roams the ambient soundscape, another can be engaged with detailed narrative content around a specific exhibit. Although this latter function is possible within a distributed, and carefully curated, speaker-based system, the greater degree to which the soundscape can be curated in audio augmented reality eliminates the problems of any unwanted bleeding and over-lapping of sources.

These advantages should be considered not just in conjunction with the practical and economic advantages of deploying a mobile-based audio augmented reality solution, but also in relation to the creation of a sonic atmosphere that can help communicate understanding of a subject, connect the listener to the subject and gallery environment and function as a catalyst for knowledge exchange.

*Curating with audio augmented objects*

Within the gallery, both the object-first and audio-first approaches can be put to work. One can see how this could be used to curate and design visitor journeys through an exhibition, or a collection of artefacts by triggering sound sources at certain times in certain locations, or in relation to other objects. Also possible, would be advertising the location of other objects in relation to the one you are currently viewing, associated objects that work well together sonically as well was contextually, guiding and suggesting potential trajectories to the listener. Such approaches build upon Zimmerman & Lorenz's concept of the attractor sound (Zimmermann and Lorenz, 2008), a feature of the LISTEN system that recommends additional artworks, via emitted and localized virtual sound sources, to users of the system based on adaptive and personalised recommendations. Giving objects within these cultural spaces the ability to audibly communicate to visitors beyond line-of-sight has the ability to provide great potential, and significant challenges, for the designers and curators of such spaces, spaces where the visual has maintained primacy from architectural design through to curatorial decision making for centuries (Connor, 2011), and therefore constitutes a new design space and way of *seeing* within such environments.

    Additionally, regarding the authorship and curation of trajectories through an exhibition space, such an approach places the object in the role of both waypoint and content, making an object potentially a functional, and a thematic, part of the system. This is considered in relation to a definition of an audio event that 'tells, or supports the telling of the narrative' as being thematic, and an audio event that 'supports participants in their navigation or comprehension of the experience' as being a functional one (Hazzard *et al.*, 2017). In addition to being functional instruments that guide and suggest exploratory trajectories through exhibition space, the object-first and audio-first

approaches can be interchanged within the context of a single exhibition depending on what curators want people to attend to.

Finally, we can draw some similarities between the interactional phases uncovered during *The McMichael Experiment*, especially the move from the *exploration* phase through to the *focused listening* and *second-level focussed listening* phases, with the modes of listening outlined by Pierre Schaeffer (1966). Just as Schaeffer's *listening modes* describe a listening journey from perception through to comprehension, the interactional phases outlined here describe a similar trajectory. Likewise, Barry Truax's *Listening-In-Search* and *Listening-In-Readiness* modes of listening (Truax, 2001) closely resemble the *investigative* and *focussed-listening* phases identified here. What perhaps the uncovered interactional phases of listening contribute within this article is that, in combination with the other findings, we are presented with a means by which listeners' trajectories through these modes can be designed.

### *Interfacing with history*

With the experiences presented here, accessing the audio archival content is achieved by using the body like a tuning dial on an analogue radio set, allowing the visitor to find clear signals within the soundscape. In relation to Mortensen and Vestergaard's similar approach (Mortensen and Vestergaard, 2014), Bijsterveld (2015) describes this as a 'highly original framing of the exhibition sounds' and one where 'the exhibition space itself mimicked the technology behind the sounds that were the topic of the exhibition'. One could also argue that this act of embodied, interactive tuning constitutes a physical contextualisation of the virtual digital archive content.

Within Mortensen and Vestergaard's installation this physical contextualisation is extended through the construction of *listening situations*, where the settings of the original physical listening environment associated with the different pieces of audio content (an armchair for content programmed in the evening, a car seat for drivetime content and a bedroom for teenage content) are reconstructed within the gallery space. Again, we see an exploration into how the material can be used to promote and focus engagement with the immaterial, a mixed-reality exercise in the contextualisation of the virtual with the physical.

Mortensen and Vestergaard's approach is largely justified by an understanding that learning associated with immersion is experience driven (Mortensen and Vestergaard, 2014). As such, the authors anticipated visitor learning outcomes to include experiencing situations, feelings and memories, not hard facts. As with the experiences presented here, their approach bears fruit in the form of positive participant feedback in relation to awareness, interest, engagement and the evocation of memories associated with the audio archive content used within the exhibition.

Understanding a little more about the dual nature of the interactional experience and character of the audio augmented object leads to some interesting considerations regarding their specific and potential function as interfaces to audio archival content. What appears to be of primary significance here is that, by augmenting a historical museum artefact with contemporaneous audio archival content, a mutual and functional contextualisation of both these components occurs; the artefact functions as an interface for the audio and provides it with physical presence and historical context, and the audio functions as an interface for the object and provides it with additional physical presence and historical context. Furthermore, this factor presents itself as a way in which obsolete and inanimate museum objects can once again become experiential objects, as well as interfaces to audio archives.

**Conclusions**

This article has outlined the characteristics, experiential qualities and functional attributes of audio augmented objects within exhibition spaces. By way of a general conclusion, it is proposed that audio augmented objects are capable of altering the perception of physical reality within museum and gallery environments, provoking memories and advertising the presence of exhibits beyond visitors' line-of-sight. Using these properties, designers of audio augmented object experiences can harness this functionality to, for example, reframe physical environments, collections of objects or, indeed, both. Such functionality, it would seem, is highly desirable for museum and gallery curators wishing to re-curate existing collections of art and artefacts while foregoing extensive physical alterations to existing displays and gallery layouts. Through the creative and contextualised design of interactional environments and the use of embodied interactions that mimic the exhibited technologies within these experiences, it is also shown that visitor exploration and engagement with exhibition spaces can be enhanced. It also appears that the agency afforded to visitors which enables them to self-compose their accompanying audio experience to an exhibition is an integral component to this engagement, and the act of exploration itself is an enjoyable activity that can be further exploited through design. Furthermore, audio augmented objects within gallery spaces provide effective opportunities for exhibiting sound, providing interfaces for digital archival audio content and engaging visitors with *silenced* exhibits.

**Limitations and future work**

It should be noted that the work presented in this article has its limitations. Firstly, the limited number of participants within the two studies should be acknowledged. Secondly, though representative across the adult age range, the studies did not involve children. This seems important, especially when considering the museum environment as a place that family groups visit. In this respect, it would seem prudent that further research within this context includes children and family groups as participants.

Additionally, it is envisaged that there are opportunities for focused, laboratory-based studies arising from the work presented here. Such studies could focus upon users' abilities to localise and engage with different types of audio content and physical objects used to construct audio augmented objects. Such studies could also make use of application logs and heat maps to better determine the types of audio content and physical object combinations that best aid exploration, attraction and engagement. Such studies could also be used to determine exactly how far acoustic realism within such experiences can be manipulated, while still maintaining the perception of reality.

**Acknowledgments**

The author would like to thank James Mansell and Annie Jameson for their work and support in helping to realise *The McMichael Experiment* and Aleks Kolkowski, kitt price and Christine Ferguson for their work and support in helping to realise *The Laboratory of Psychical Research* AR experience. The author would also like to acknowledge their current research funding: "XTREME - Mixed Reality Environment for Immersive Experience of Art and Culture" EU HORIZON grant EU HORIZONCL42023-HUMAN-01-CNECT XTREME (grant no. 101136006).

**Declaration of interests**

The author reports there are no competing interests to declare.